\documentclass[a4paper,11pt]{article}
\pdfoutput=1 

\usepackage{jheppub} 

\usepackage[T1]{fontenc} 

\title{ Recent advances in accelerated discovery through machine learning and statistical inference}

\author[a,*]{Ryan B. Jadrich}
\author[a,*]{Beth A. Lindquist}
\author[a,b,1]{Thomas M. Truskett\note{Corresponding author.}}

\affiliation[a]{McKetta Department of Chemical Engineering, University of Texas at Austin, Austin, Texas 78712, USA}
\affiliation[b]{Department of Physics, University of Texas at Austin, Austin, Texas 78712, USA}
\affiliation[*]{These authors contributed equally to this review.}

\emailAdd{truskett@che.utexas.edu}

\abstract{Recent applications of machine learning and statistical inference provide case studies demonstrating how such approaches can accelerate the discovery process in physical chemistry and related fields. Examples discussed in this review include the introduction of automated approaches to systematically improve experimental design, increase the efficiency of computationally expensive molecular simulations, facilitate construction of predictive models for complex biological processes, and discover interparticle potentials that lead to materials which meet specified design goals. A common theme is the synergy between experiment and computation enabled by such approaches.}

\begin{document} 
\maketitle
\flushbottom

\section{Introduction}

Modern computation is integral to research in the physical sciences. In the present ``Age of Big Data'',~\cite{big_data,big_data_2,petabytes_high_energy} there are increasing opportunities for computational research to facilitate scientific discovery via Artificial Intelligence (AI). Historically, much of the academic work in AI was beset by disappointment, with early promises left unfulfilled, culminating in a steep decline in funding for AI research in the 1970s called the ``AI winter''.~\cite{lighthill,ai_book,tumultuous_search} In some ways, the overly optimistic predictions of the 1960s have still not come to pass (e.g., the sentient computer HAL 9000 from Arthur C. Clarke's \emph{Space Odyssey}).~\cite{space_odyssey} However, there have also been compelling successes, including the victory of Google's ``AlphaGo'' over a professional human opponent, that have been fueled by a resurgence of interest in AI.~\cite{alpha_go_bbc,alpha_go_science,alpha_go_intro} Current work in the field is often guided by a practical data-driven goal:  the ability to automatically extract specific insights from (or make predictions based on) large data sets.~\cite{ai_book,bayesian_reasoning,statistical_learning} Thus, while sentient robots may not exist, more subtle achievements have been realized including the introduction of powerful machine learning (ML) and statistical inference (SI) approaches. Such methodologies have the potential to be exploited in the physical sciences in order to 1) extract key insights, 2) make research processes more systematic, and 3) increase research efficiency. This review details recent progress towards this end. 

Both ML and SI take as input an array of feature variables and attempt to fit some model to a known ``training'' set of data. Fitting the model involves ``learning'' the values of the tunable model parameters via a systematic optimization technique.~\cite{bayesian_reasoning,statistical_learning} The distinction between ML and SI that we adopt in this work lies in how the model is used. Typically, ML employs powerful black-box models to accurately predict some unknown quantity of interest that is believed to be dependent on the features. In model fitting, the training data is supplemented with prior observations of the relationship between features and the unknown quantity. For SI, the goal is typically to model the underlying process by which some set of features is generated. The process that generates the observables of interest is assumed to be representable by a probability distribution with parameters that are ``learned'' or ``inferred''.

ML and SI are not wholly new to the physical sciences. For example, widely used regression analysis methods are fundamentally ML methods. More recently though, increasingly advanced ML and SI strategies are being leveraged to solve a variety of challenging problems. In this review, we first describe the use of ML predictions to expedite the iterative experimental design process of synthesis and measurement for maximizing a quantifiable design goal.~\cite{alloy_design,another_alloy_design,phase_boundary_design,using_uncertainties} By using ML-based methods, the results of previous experiments can be used to predict the most promising new sample to synthesize or new experiment to perform. In another study, ML is leveraged to expedite the computation of quantum mechanical (QM) forces for use in molecular dynamics (MD) simulations.~\cite{molecular_dynamics_1,molecular_dynamics_2,force_fields} Prior QM calculations are used to create a predictive ML model relating configurations to QM forces. 

Applications of SI in physical chemistry have also emerged to address a variety of challenges. First, we review the recent application of probabilistic statistical mechanical approaches for modeling complex biological phenomena. Application to virus models yields insight into how various residues interact, which could potentially expedite vaccine design.~\cite{hiv_model,hiv_model_confirmation,hepc_model,error_catastrophe} Similarly, SI methods have been employed to extract intra-cellular interaction networks on the basis of DNA microarray data.~\cite{gene_interactions} Finally, the review ends with discussion of two new inverse design strategies based on SI for the systematic discovery of materials which either exhibit complex self-assembly behavior~\cite{clusters_pores_crystals,crystals,pores,many_pores,clusters} or possess a desired material property~\cite{design_engines}. Both methods learn the requisite interactions via relatively simple optimization schemes.

\section{Acceleration of conventional optimization and simulation schemes} 
\label{sec:accelerate}

A straightforward application of ML methodology is the prediction of a quantity of interest, $F$, as a function of known input conditions, $\boldsymbol{X}\equiv\{x_{i}\}$, called features; here, $\{...\}$ indicates a set or vector of values of some unspecified size. Generally, $F$ is difficult or costly to evaluate.~\cite{bayesian_reasoning,statistical_learning,bayes_thesis} Therefore, accurate predictions are useful to save time and resources. For so-called supervised ML strategies, a predictive model, $F(\boldsymbol{X}|\boldsymbol{\theta})$, is fit via flexible parameters $\boldsymbol{\theta}\equiv\{\theta_{i}\}$ to the prior observations of the correspondence between $\boldsymbol{X}$ and $F$, i.e., the training data, $\mathcal{D}\equiv \big\{ \big(F^{(i)},\boldsymbol{X}^{(i)}\big) \big\}$. The model allows for prediction of $F$ at $\boldsymbol{X}$ not included in the training set. Uncertainties associated with the predictions for $F$ can be also obtained to yield a probabilistic model for the data, $P(F|\boldsymbol{X},\boldsymbol{\theta})$. The uncertainties can either be directly obtained via ML approaches (e.g., Gaussian process models)~\cite{bayesian_reasoning,statistical_learning,bayes_thesis,bayes_opt_kriging_believer}, or computed via bootstrapping~\cite{statistical_learning}, i.e., re-sampling the data to generate new model fits that probe the sensitivity of the model to fluctuations in $\mathcal{D}$. For either approach, additional sampling both improves the prediction and reduces the uncertainty in $P(F|\boldsymbol{X},\boldsymbol{\theta})$.

Supervised ML methods essentially amount to interpolation strategies, each with different underlying assumptions regarding function form, smoothness and possibly the degree of interaction among the features. For example, the popular Gaussian process model requires a ``kernel'' that governs how correlations between two points in feature space decay, the choice of which influences the smoothness and uncertainty of the predicted function.~\cite{bayesian_reasoning,statistical_learning,bayes_thesis,bayes_opt_kriging_believer} Another important consideration for ML is the construction of the features. It is oftentimes advantageous to purposefully engineer features to reflect the underlying physics of the problem, instead of requiring the model to learn these relationships via sampling. Feature engineering is non-unique and requires careful consideration of the problem at hand but can ultimately lead to better convergence of the model.~\cite{bayesian_reasoning,statistical_learning} Examples of some feature choices will be discussed in the next two subsections, which are organized as follows. In Sect.~\ref{subsec:bo}, we consider how prediction of the figure of merit (or fitness) associated with a set of tunable conditions can be used to accelerate certain optimization problems. In Sect.~\ref{subsec:qmmd}, we discuss the use of ML techniques to approximate QM forces for use in MD simulations.  

\subsection{Optimization}
\label{subsec:bo}

There are myriad optimization schemes that can be used to maximize a desired property, termed the figure of merit (FOM), by tuning a set of adjustable parameters, $\boldsymbol{\theta}$.~\cite{bayesian_reasoning,statistical_learning,numerical_recipes} Most require the use of derivatives with respect to $\boldsymbol{\theta}$. However, there are also many cases where derivatives are not easily obtained, e.g., when analytical derivatives are not available (or do not exist) and there are too many adjustable parameters for numerical differentiation to be practical. For such cases, heuristic optimization strategies such as simulated annealing or genetic algorithms have found widespread use.~\cite{numerical_recipes} In these methods, updates to parameters are governed by a tension between increasing the FOM and promoting a wider exploration of parameter space through stochastic processes. As a result, a large number of FOM evaluations can be required to arrive at a solution, limiting the practicality of such strategies, especially for FOM evaluations that require expensive computations or experiments. To handle such cases, ML techniques can be employed in order to generate parameter updates that are informed by prior FOM evaluations as opposed to stochastic processes that only consider possible changes to the current FOM.

Bayesian optimization (BO) represents one such approach, whereby a probability distribution, $P(F|\boldsymbol{X},\boldsymbol{\theta})$, is constructed to model the true underlying FOM function or landscape, $F_{0}\equiv F_{0}(\boldsymbol{X})$, by fitting to prior observations (i.e., $\mathcal{D}$ above) via the tunable model parameters ($\boldsymbol{\theta}$).~\cite{bayes_thesis,bayes_opt_kriging_believer} $P(F|\boldsymbol{X},\boldsymbol{\theta})$ provides information on regions in parameter space that are 1) highly likely (determined via prior FOM evaluations) to possess high $F$ values and 2) those of low certainty but large positive uncertainty in $F$. The former can be \emph{exploited} to further refine the best-known parameters, and the latter represent profitable areas for \emph{exploration}--allowing the optimization to progress to other maxima as needed. As more observations are included, progressively better predictions for $F_{0}(\boldsymbol{X})$ are generated.   

After construction of $P(F|\boldsymbol{X},\boldsymbol{\theta})$, the model must be translated into a prediction for a set of parameters $\boldsymbol{\theta}$ for the next optimization step.~\cite{bayes_thesis,bayes_opt_kriging_believer} Because this mapping is non-unique, multiple options exist for this task, each providing a particular balance of exploitation and exploration. One such choice for a recommender function, $R(\boldsymbol{X}|\boldsymbol{\theta})$, is called Expected Improvement
\begin{equation} \label{eqn:decision_function}
R(\boldsymbol{X}|\boldsymbol{\theta})\equiv \langle \max(0,F-F^{*})\rangle_{P(F|\boldsymbol{X},\boldsymbol{\theta})}
\end{equation}
where $F^{*}$ is often taken as the best FOM evaluation at any prior $\boldsymbol{X}$.~\cite{bayes_opt_kriging_believer} Only regions of uncertainty that are better than the current optimal set of parameters are included, i.e., uncertainty in the negative direction is not penalized. Including the uncertainty associated with the predictions in the recommender function has been shown to enhance the optimization efficacy.~\cite{using_uncertainties} As the optimization progresses, uncertainty will decrease in regions of parameter space that are sampled, perhaps directing the optimization to explore areas with greater uncertainty. 

BO can be generalized to suggest the most promising ``batch'' of samples, of size $q$.~\cite{bayes_thesis,bayes_opt_kriging_believer} In computational research, $q$ may be the number of parallel processes that can be run, while in experiment, $q$ might be governed by the sample auto-feeder size, for instance. One generalization adopted in the work below is called the Kriging believer algorithm, where the model is iteratively updated via predictions for $F$ at suggested $\boldsymbol{X}$ within the same batch (i.e., ``believe'' the model). Once actual measurements are taken, the measured FOM values replace the predicted ones in the data set. 

One recent study integrated BO into the experimental design of a NiTi-based shape memory alloy (SMA) with minimized thermal hysteresis.~\cite{alloy_design} SMAs exhibit shape memory and super-elasticity as the result of an underlying high-temperature austenite and low-temperature martensite phase transition. The temperature at which the transition occurs depends on whether the material is being heated or cooled, yielding a hysteresis temperature gap, $\Delta T$. Larger values for $\Delta T$ result in greater performance degradation--motivating the chosen FOM definition: $F=-\Delta T$.  Optimization targeted an alloy of the form $\text{Ni}_{50-x-y-z}\text{Ti}_{50}\text{Cu}_{x}\text{Fe}_{y}\text{Pd}_{z}$ where $x$, $y$, and $z$ are doping elements allowed in increments of $0.1\%$ within the constraints $50-x-y-z\geq 30\%$, $x\leq 20\%$, $y\leq 5\%$ and $z\leq 20\%$--yielding 797,504 possible combinations. The combinatorial challenge combined with the expensive FOM evaluation (experimental synthesis) called for a Bayesian approach. 

\begin{figure}[h]
\centering
\includegraphics[width=5.5in]{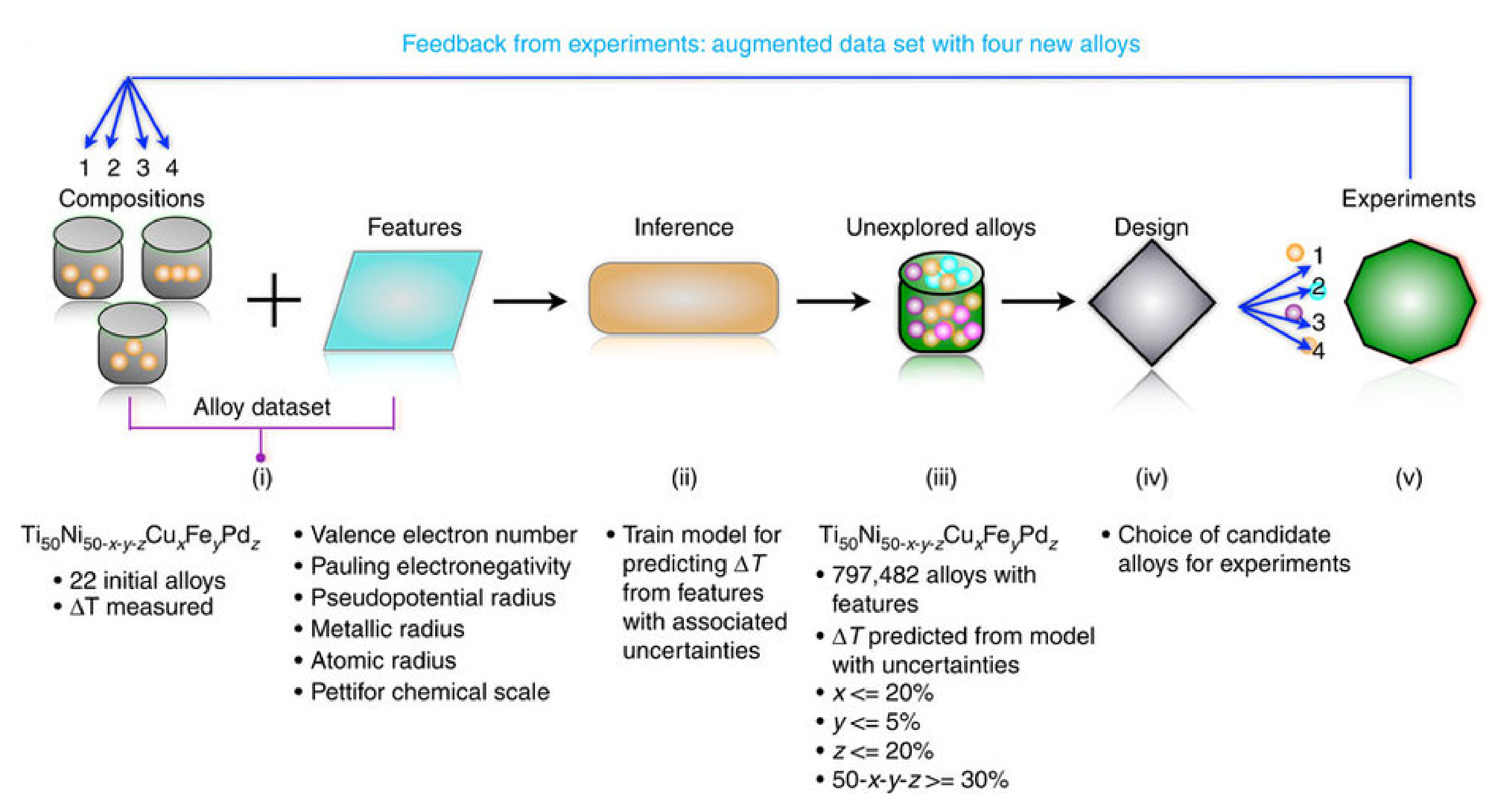}
\caption{Iterative design strategy for minimizing the hysteresis of an alloy. Beginning with a set of training samples to initialize the inference procedure, new samples that hold promise are predicted, synthesized, and the associated hysteresis measured; the new results are then used to augment the training set before beginning the cycle anew. Figure reproduced from~\cite{alloy_design}.}
\label{fig:alloy_opt_loop}
\end{figure}

BO, like all ML approaches, requires defining the features ($\boldsymbol{X}$) to be used as input to the ML algorithm. While $x$, $y$ and $z$ are the tunable experimental parameters, their effect on $\Delta T$ is mediated by their impact on various atomic properties such as valence electron number. Therefore, the authors represented features via weighted fractions (e.g., $f_{x}=x/(x+y+z)$) of the atomic properties listed in the second column of Fig.~\ref{fig:alloy_opt_loop}. Each property is a coarse-grained electronic structure measure known to influence the phase transition temperatures and hysteresis. While the ML model might eventually learn these relationships with sufficient sampling, directly encoding them is likely to improve the convergence rate of the model. 

The overall iterative design strategy for alloy discovery is summarized in Fig.~\ref{fig:alloy_opt_loop}. The initial $\mathcal{D}$ was composed of 22 samples, which for each, differential scanning calorimetry measured $\Delta T$. From these samples, the features were computed and used as input to the probabilistic model, from which a batch of new unexplored alloys with the highest expected improvements were drawn. The FOMs ($-\Delta T$) associated with the new parameter combinations were experimentally determined, and these data points were then used to update the probabilistic model, from which new, more accurate, parameter combinations were extracted in an iterative fashion. After only nine iterations of $q=4$ batch updates, 14 out of the 36 total new alloys posses a $\Delta T<3.15K$, which was the best of the initial 22 training samples. The outcome of the BO was a $42\%$ improvement over the best initial training sample; the alloy $\text{Ti}_{50.0}\text{Ni}_{46.7}\text{Cu}_{0.8}\text{Fe}_{2.3}\text{Pd}_{0.2}$ with $\Delta T=1.84K$ was discovered. 

In conclusion, BO is a promising avenue for assisting in experimental (or computational) design. The approach is general and can be adapted to accommodate other scenarios. For example, the same authors recently leveraged BO for the accelerated search of BaTiO$_{3}$-based piezoelectrics to minimize the curvature of a phase boundary.~\cite{phase_boundary_design} Utilizing various descriptive features, prior phase diagrams, and the predicted functional form (from Landau theory) for the phase boundaries, materials with better piezoelectric properties than any in the initial training set were discovered. Beyond hard materials, the field of soft materials design could benefit from BO as 1) database knowledge is limited, and 2) experiments are often highly customized and expensive. Conveniently, many simplified theories exist for soft materials,~\cite{soft_matter_book_1,soft_matter_book_2} the predictions of which can be incorporated into the optimization as features to embed prior knowledge and improve model convergence.

\subsection{Molecular Simulation}
\label{subsec:qmmd}

MD simulations require many interparticle force evaluations especially for systems with large numbers of particles, motivating the development of classical force fields. Such models invoke a selected functional form (e.g., a quadratic function for a chemical bond) with flexible model parameters (e.g., the associated spring constant) that are fit to reference data, which might be from either QM calculations or experiments. However, classical force fields are not always appropriate; an archetypal failure is that a harmonic spring cannot model bond-breaking processes. Ideally for such cases, QM data would be used to determine the forces for MD simulations; however, QM calculations that are sufficiently accurate to produce meaningful MD simulations are often too computationally expensive.

A ML model that is able to predict QM forces for a given particle arrangement would significantly reduce the computational expense of MD simulations with QM forces. Such a strategy would be particularly amenable to standard ML approaches due to the high degree of structural similarity between many of the configurations that arise in a given MD simulation. In general, proper sampling in MD studies results in an ensemble of configurations possessing similar structural correlations. Therefore, ML models can be employed to learn the complex relationships between QM forces and the configurations for some initial set of structures that initialize the model (i.e., the training data); the resulting model can then be used to \emph{predict} QM forces.~\cite{molecular_dynamics_1,molecular_dynamics_2,force_fields} 

\begin{figure}[h]
\centering
\includegraphics[width=5.5in,keepaspectratio]{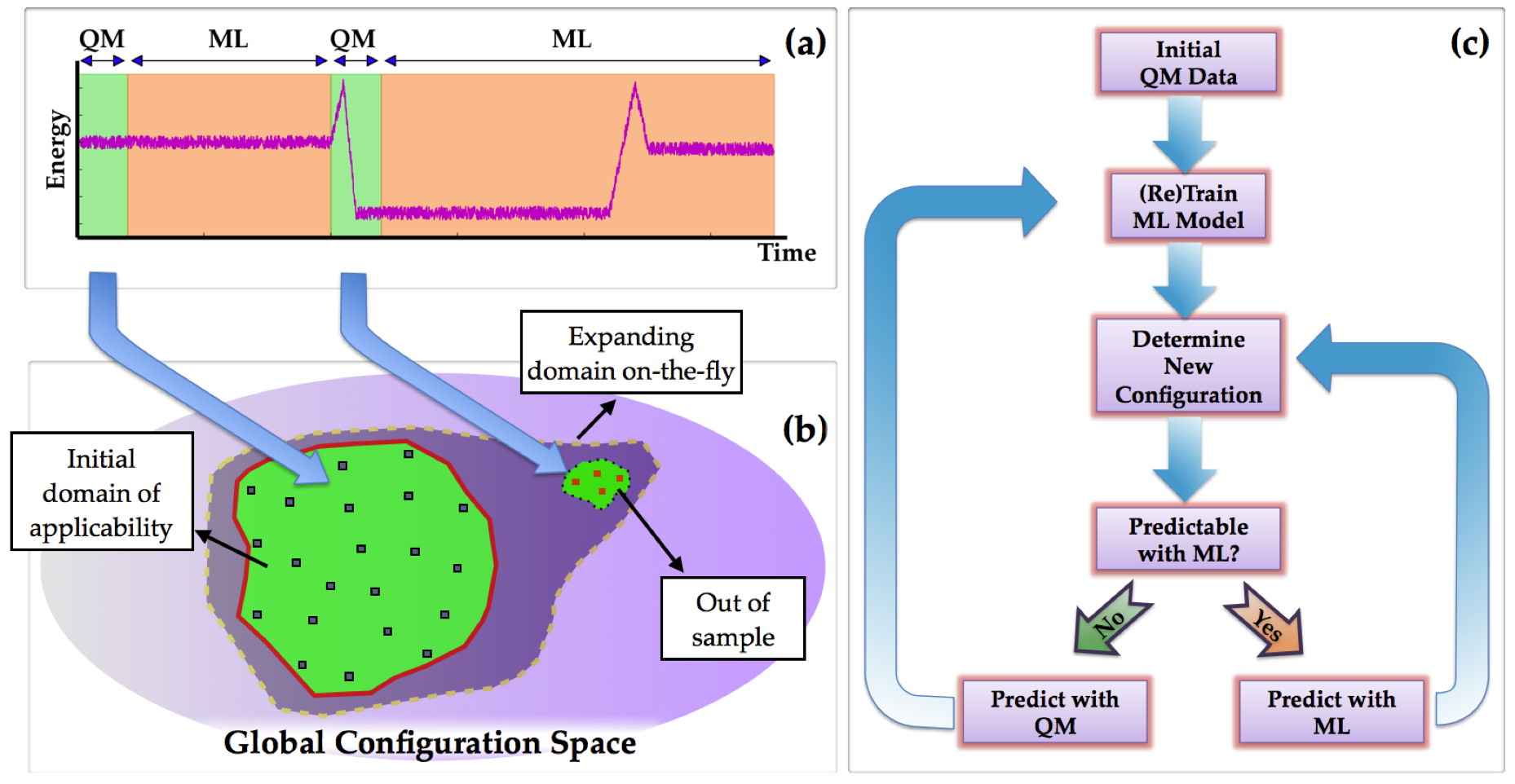}
\caption{Scheme to augment QM MD simulations with machine learning. (a) A cartoon of an MD energy trace demonstrating the switching behavior between ML predictions and full QM calculations. (b) Qualitative viewpoint of the configuration space sampled during the MD run and the use of QM calculations to expand the knowledge base on-the-fly as uncertain configurations are encountered. (c) Flowchart for the hybrid QM MD scheme. Figure reproduced with permission from~\cite{molecular_dynamics_1}.}
\label{fgr:QM_opt_cycle}
\end{figure}

The above scheme is summarized in Fig.~\ref{fgr:QM_opt_cycle}. In Fig.~\ref{fgr:QM_opt_cycle}a, the energy trace of a sample trajectory is shown; green regions indicate steps for which QM calculations are needed. In this case, the model is initialized using the early steps of the trajectory until sufficient sampling occurs such that an accurate predictive model can be constructed, at which point ML is used to predict the QM forces (the peach colored region). Later, a large fluctuation--signaled by the jump in the energy--produces a configuration that is outside of scope of the training data; see Fig.~\ref{fgr:QM_opt_cycle}b. At this point, explicit QM calculations are performed to augment the training data until this new area is ``learned'' sufficiently well that the ML predictor can be used again. In the future, large fluctuations that span these two domains can be treated using the augmented ML model. Fig.~\ref{fgr:QM_opt_cycle}c depicts this procedure as a flowchart. 

The above approach was employed to perform MD simulations of atomic Aluminum (Al) using density functional theory (DFT) to compute the QM forces for the training data.~\cite{molecular_dynamics_1,molecular_dynamics_2} Intuitively, the features--the known quantities from which the forces are predicted--must depend on the environment of the selected particle, in this case the other Al atoms in the simulation. In order to explicitly encode the physics of the problem, atomic coordinates were mathematically transformed into features that are invariant to any rotation, translation or permutation of identical particles. Additionally, the forces on a given Al atom should most strongly depend on its closest neighbors, with more distant Al atoms contributing less strongly. Therefore, both a damping function and a cutoff distance were included in the features to encode the primarily local effects of neighboring particles on the forces.

Inputting these engineered features into a standard ML model enabled quick and accurate force prediction. The errors in ML forces were found to be within the expected accuracy of DFT itself while reducing the computational expense by multiple orders of magnitude with respect to DFT. Such approaches seem promising and may enable QM MD simulations to reach time and length scales not previously believed to be possible. However, further research is needed to address certain open questions, e.g., how does one detect when the model is outside its regime of predictability? The approach adopted by the authors assumed any feature vector lying outside a hypercube encompassing the training set needed explicit DFT computation. While certainly true, points within the sample region could also warrant explicit computation. One possibility includes modeling not only the expected value but also uncertainty (as done in Bayesian optimization). If the uncertainty exceeds some threshold, an QM computation could be performed instead. Some recent work based on the distance between the new feature set and the closest neighbor in the training data shows promise.~\cite{force_fields}

\section{Modeling biology with statistical mechanics} 
\label{sec:bio}

In Sect.~\ref{sec:accelerate}, we reviewed use of ML techniques that prioritize the accuracy of predictions, in general leaving the relationships between the features and the predicted values obscured. In this Section, we consider applications of SI where a primary goal of the modeling is to understand the underlying physics that gives rise to a set of observables, potentially in conjunction with making accurate predictions for a quantity of interest. In the following studies, SI methods are used to construct a model, $P(\boldsymbol{X})$, for some underlying probability distribution, $P_{0}(\boldsymbol{X})$, where $P_{0}(\boldsymbol{X})$ is generally unknown but can be sampled from, e.g., via experimentally obtained statistics. 

SI methods have been successfully applied to systems of biological relevance. Here we review two such cases involving 1) viral sequences~\cite{hiv_model,hiv_model_confirmation,hepc_model,error_catastrophe} and 2) gene expression data~\cite{gene_interactions}. A common challenge is the large number of features that are characteristic to the problem, namely, $\boldsymbol{X}=[\boldsymbol{x}_{1}, \boldsymbol{x}_{2},...,\boldsymbol{x}_{q}]$ where $q$ is large. The number of samples required to adequately sample a hyperspace grows exponentially with $q$, colloquially referred to as the ``curse of dimensionality''.~\cite{statistical_learning} In such situations, where it is only possible to sample a small fraction of the possible feature combinations, the data will necessarily be sparse. The problem then is to formulate a model that can take such sparse sampling and make probability predictions for any $\boldsymbol{X}$. 

While statistically significant sampling of the full coordinate space, $\boldsymbol{X}$, is challenging in high dimension, it is generally possible to obtain good statistics on quantities that depend on only small subset of coordinates--providing some constraints on $P(\boldsymbol{X})$. For example, statistically significant empirical calculation of one- and two-point probability distributions, $P_{0}^{(i)}(\boldsymbol{x}_{i})$ and $P_{0}^{(i,j)}(\boldsymbol{x}_{i}, \boldsymbol{x}_{j})$ is often possible. (For example, the radial distribution function, $g(r)$, can be reliably computed from an MD simulation that samples only a minuscule fraction of possible particle arrangements.) Other quantities such as the mean $\langle \boldsymbol{x}_{i} \rangle_{P_{0}(\boldsymbol{X})}$ and co-variance $\langle \boldsymbol{x}_{i}^{T}\boldsymbol{x}_{j} \rangle_{P_{0}(\boldsymbol{X})}$ may also be reliably estimated. However, enforcing $P(\boldsymbol{X})$ to replicate lower-order correlations is an under-determined problem. A potentially infinite number of $P(\boldsymbol{X})$ can reproduce such lower-order statistics. The challenge is to decide which of the innumerable possibilities for $P(\boldsymbol{X})$ is ``best''.

Generally speaking, the best model choice given limited information is the one that is the least predictive (or least ``informative'') while reproducing the statistics that are known for $P(\boldsymbol{X})$.~\cite{jaynes} For example, two of the infinite probability distributions for a six-sided die that are known to yield an average face value of 3.5 are 1) a fair six-sided die (i.e., each face ($f=1-6$) is equally likely, or $P(f)=1/6$ for all $f$), and 2) a die that only turns up a 3 or a 4 in equal proportion ($P(3)=P(4)=1/2, P(f)=0$ where $f=1,2,5,6$). Intuitively, the $P(f)$ for a fair die contains the least information possible--the distribution is maximally ``smeared out'' across possible states; by contrast, the latter distribution \emph{predicts} that the face value of the roll will either be a $3$ or a $4$. For a more complex constraint though, the ``best'' $P(\boldsymbol{X})$ may not so intuitive. Therefore, we require a measure quantifying the information content of a distribution $P(\boldsymbol{X})$.

The maximum entropy principle states that the best model of the many that meet the desired constraints is the one that maximizes the Shannon entropy~\cite{jaynes,shannon,max_cal,generalized_max_ent,spin_glasses_and_max_ent}
\begin{equation} \label{eqn:shannon_entropy}H[P(\boldsymbol{X})]\equiv -\sum_{\boldsymbol{X}}^{} P(\boldsymbol{X})\text{ln}P(\boldsymbol{X})=-\langle\text{ln}P(\textbf{X})\rangle_{P(\textbf{X})}
\end{equation}
where $\sum_{\boldsymbol{X}}^{}$ is shorthand for a summation over all possible $\boldsymbol{X}$. In practice, maximum entropy is typically carried out as a two-step protocol. First, Eqn.~\ref{eqn:shannon_entropy} is analytically maximized subject to the constraints via the technique of Lagrange multipliers. This provides a functional form for the distribution that is parametrized by the undetermined Lagrange multipliers, which are then solved for given the known constraints.

\subsection{Viral Fitness Landscapes}

Preventative vaccination represents the most realistic way of combating viruses such as Hepatitis C virus (HCV) and human immunodeficiency virus (HIV-I). However, many viruses have high sequence mutability, providing pathways to escape from vaccine-induced immunological pressure. Therefore, effective vaccines might need to account for the entirety of the vast combinatorial ``sequence space'' available to the virus. Unfortunately, relatively limited amounts of statistical data (from experimental observations of virus sequences) are available in comparison to the high dimensionality of the problem. Therefore, SI strategies based on maximizing the Shannon entropy are appropriate for this task.

The ``fitness'' associated with a specific sequence is determined by its ability to replicate. While this quantity can be measured by \emph{in vitro} studies, existing experimental data of this type are sparse. Instead, clinical databases that quantify the frequency of given sequences are used as a proxy for the fitness; these data span a much larger range of sequences. The underlying, physically reasonable, assumption is therefore that ``fitter'' sequences are more prevalent in the real world~\cite{hiv_model_confirmation}. The goal of a vaccine is to induce immunological pressure on fit sequences, thus driving the viral population into unfit sequence space where it is minimally damaging to the infected host. 

Both HCV and HIV-1 models followed from maximization of the Shannon entropy (Eqn.~\ref{eqn:shannon_entropy}) subject to reproducing the empirical one- and two-point sequence distributions $P_{0}^{(i)}(s_{i})$ and $P_{0}^{(i,j)}(s_{i}, s_{j})$, respectively, where $s_{i}$ is the amino acid identity at the $i^{th}$ location in the considered sequence ($\boldsymbol{S}\equiv[s_{1},s_{2},...,s_{N_{S}}]$).~\cite{hiv_model,hiv_model_confirmation,hepc_model,error_catastrophe} The model takes on the following statistical mechanical form:
\begin{equation} \label{eqn:max_ent_one_two_constraint}
\begin{split}
 & P(\boldsymbol{S}|\boldsymbol{K})=\exp[-E(\textbf{S}|\boldsymbol{K})]/Z(\boldsymbol{K}) \\
 & Z(\boldsymbol{K})\equiv \sum_{\boldsymbol{S}}^{}\exp[-E(\textbf{S}|\boldsymbol{K})] \\
 & E(\textbf{S}|\boldsymbol{K})\equiv \sum_{i=1}^{N_{S}}K^{(i)}(s_{i}) + \sum_{i=1}^{N_{S}}\sum_{j=i+1}^{N_{S}}K^{(i,j)}(s_{i}, s_{j}) 
\end{split}
\end{equation}
where $\boldsymbol{K}$ is the combined set of all one- and two-point dimensionless ``energy'' functions, $K^{(i)}(s_{i})$ and $K^{(i,j)}(s_{i}, s_{j})$, that need to be determined such that $P^{(i)}(s_{i})=P_{0}^{(i)}(s_{i})$ and $P^{(i,j)}(s_{i}, s_{j})=P_{0}^{(i,j)}(s_{i}, s_{j})$. Formally, this model is identical to the infinite range Potts spin model in statistical physics.~\cite{potts_model} It would be possible to match higher-order correlations as well via a multi-body energy expansion; however, the sample size requirements for adequate statistical estimation of correlations increase exponentially with the dimensionality of the correlation quantity. Attempting to exactly reproduce higher-order correlations that have not been adequately sampled will generally result in over-fitting and therefore reduced transferability to data that is not included in the modeling step. Nonetheless, the above models based on one- and two-body correlations constructed for viruses were found to also reproduce three-body correlations, even though these were not explicitly considered for modeling.~\cite{hepc_model}

\begin{figure}[h]
\centering
\includegraphics[width=5.5in,keepaspectratio]{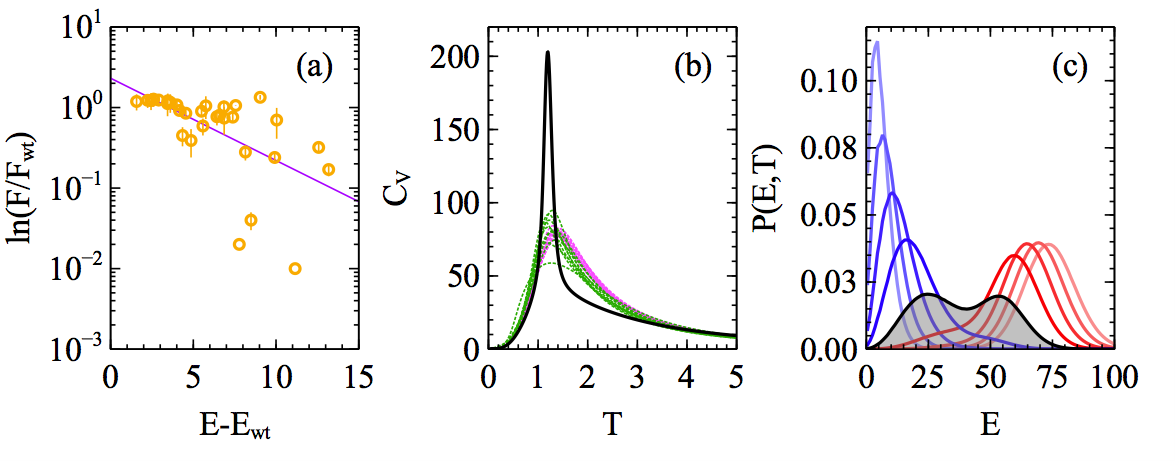}
\caption{(a) Comparison of \emph{in vitro} fitness measurements (normalized with respect to the wild-type) for 31 engineered HCV NS5B mutants to maximum entropy model predictions. Overall, qualitatively reasonable agreement is found. (b) The solid black curve is the fictive ``heat capacity'' as a function of temperature for the HIV P6 protein model, providing a thermodynamic indication of the phase transition. Other heat capacity curves are generated for random spin models not obtained by fitting to the data--indicating the phase transition is a characteristic of the virus, not the modeling approach. (c) Probability for sequences of a particular energy at a specified fictive model temperature in the P6 protein model for HIV. From left to right, $T=0.8-1.6$ in increments of $0.1$ The bi-modal character that emerges at $T=1.20$ indicates the existence of two populations (high and low energy) that coexist (equal area under the curve)--corresponding to a phase transition in a finite-sized system. Figure adapted from~\cite{hepc_model,error_catastrophe}.}
\label{fig:viral_model_validation}
\end{figure}

The models were validated by a multifaceted analysis which tested the ability of the model to 1) infer sequence fitness, 2) identify viable escape mutations for the virus, and 3) agree with known effective immune responses that target specific regions of the viral sequence. With respect to the first component, one expects measurements of the \emph{in vitro} fitness, $F$, to be related to the model via $\text{ln}F(\textbf{S}|\boldsymbol{K})\propto -E(\textbf{S}|\boldsymbol{K})$ if prevalence and fitness are proportional. As shown in Fig.~\ref{fig:viral_model_validation}a, this relation is found to be reasonably well approximated in HCV--an analogous study showed this to hold for HIV as well. Other more specific tests (such as identifying escape mutations) demonstrated similar success--validating the model and its underlying assumption of fitness and prevalence.

In addition to providing an avenue to deal with the high dimensionality of the problem, the use of a spin model for viral sequences provides the added benefit of interpretability by way of analogy to actual statistical mechanical systems. In modeling HIV-I, a phase transition was observed that might be a consequence of viral ``error catastrophe'', a well-known concept where accelerating the mutation rate of the virus to a certain tipping point eventually leads to viral extinction.~\cite{error_catastrophe} Above the transition, fatal error accumulation leads to an ensemble populated by many high energy (and therefore low fitness) strains. RNA viruses in particular are thought to have evolved to operate in close proximity to the error catastrophe, so as to be maximally mutable--enhancing the ability of viruses to escape the host's immune response--while remaining effective.~\cite{rna_error_cat} Purely theoretical models have been developed that display the error catastrophe; however, such models necessarily invoke uncontrolled assumptions about viral replication.~\cite{error_catastrophe_model} Empirical modeling of viral fitness as described above, on the other hand, is based upon clinical data that implicitly accounts for factors related to viral fitness and replication.    

By defining a fictive temperature ($T$), which corresponds to a re-scaling of the energy via $E(\boldsymbol{S})\rightarrow E(\boldsymbol{S})/T$ in Eqn.~\ref{eqn:max_ent_one_two_constraint}, a corresponding ``heat capacity'' was computed for the HIV-I fitness model. The sharp spike shown in Fig.~\ref{fig:viral_model_validation}b is suggestive of an underlying phase transition. This phase transition was interpreted as a signature of the error catastrophe because it signifies a change in the population preference for low versus high fitness sequences, as can be observed in Fig.~\ref{fig:viral_model_validation}c. Specifically, below the transition temperature ($T_{\text{ec}}=1.20$), the model virus exists as an ensemble of relatively few high fitness (i.e., low energy) strains, with the distribution narrowing and shifting to lowering energies as $T$ decreases. Conversely, the model predicts a generally broader and much higher energy distribution above $T_{\text{ec}}$. However, at $T_{\text{ec}}=1.20$, there is a co-existence between high and low fitness strains, as evidenced by the bimodality in the black curve in Fig.~\ref{fig:viral_model_validation}c. Of course, this co-existence between two ``states'' is also a well-known characteristic of a first-order phase change in systems that obey equilibrium statistical mechanics. 

This research provides the first empirical HIV model to corroborate the limited experimental reports supporting the existence of an accessible error catastrophe. It has been proposed that altering the model via tuning $T$ corresponds to varying the mutation rate: 1) $T\rightarrow\infty$ corresponds to random copying, and 2) $T\rightarrow 0$ corresponds to perfect, high-fidelity replication--a variable which could be exploited in the search for a cure. Interestingly, such effective temperatures may be reachable by mutation rate enhancing drug therapies, and some clinical trials for an HIV mutagen have already begun.~\cite{error_catastrophe_exp_1,error_catastrophe_exp_2,error_catastrophe_exp_3}

\subsection{Maximum entropy inference for complex systems}
\label{subsec:pairwise}

In the studies above, it might seem surprising that inclusion of only one- and two-body correlations is sufficient to accurately model the many-body relationship between sequence and viral fitness. However, recent work indicates that the complexity inherent to biological systems might actually contribute to the successes of pairwise models in approximating their behavior.~\cite{pairwise_sufficiency} A prior study of simplified Ising p-spin models~\cite{p_spin} indicates that the sufficiency of a pairwise model is tied to the entropy of the distribution being approximated. p-spin models describe a set of $N_{q}$ spins $\boldsymbol{Q}\equiv[q_{1},q_{2},...,q_{N_{q}}]$ with possible values $q_{i}=\{+1,-1\}$ as follows
\begin{equation} \label{eqn:p_spin_model}
\begin{split}
& P_{p}(\boldsymbol{Q}|\boldsymbol{J}_{p})=\exp[-E_{p}(\boldsymbol{Q}|\boldsymbol{J}_{p})]/Z(\boldsymbol{J}_{p}) \\
& Z_{p}(\boldsymbol{J}_{p})\equiv \sum_{\boldsymbol{Q}}^{}\exp[-E_{p}(\boldsymbol{Q}|\boldsymbol{J}_{p})] \\
& E_{p}(\boldsymbol{Q}|\boldsymbol{J}_{p})\equiv \sum_{\nu_{1}=1}^{N_{q}}\sum_{\nu_{2}=\nu_{1}+1}^{N_{q}}\cdot\cdot\cdot \sum_{\nu_{p}=\nu_{p-1}+1}^{N_{q}}       J_{\nu_{1},\nu_{2},...,\nu_{p}}q_{\nu_{1}}q_{\nu_{2}}\cdot\cdot\cdot q_{\nu_{p}}
\end{split}
\end{equation}
where $\boldsymbol{J}_{p}$ is shorthand for the set of all p-body spin couplings, $\{J_{\nu_{1},\nu_{2},...,\nu_{p}}\}$. 

\begin{figure}[h]
\centering
\includegraphics[width=5.5in]{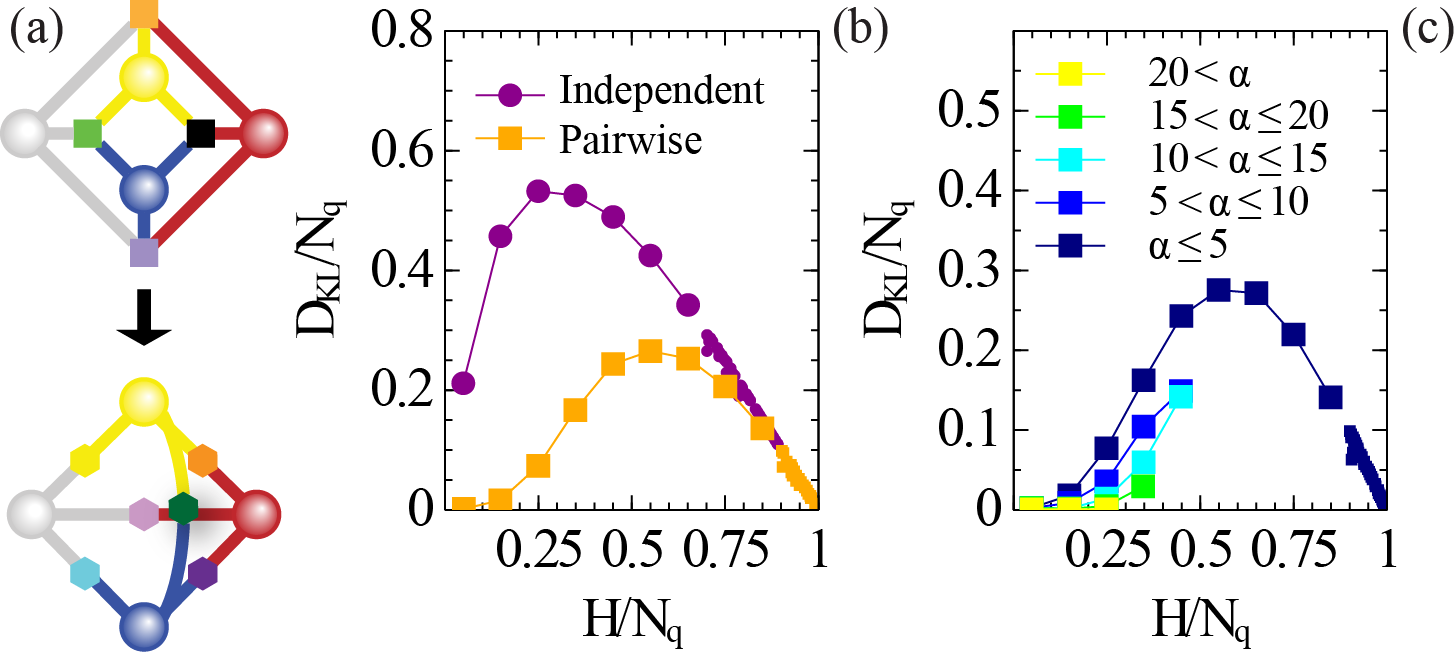}
\caption{(a) Schematic depicting the mapping of a 3-body spin model onto a pairwise spin model for four spins (the circles). The 3-body interactions are represented with squares, and the pairwise interactions are shown as hexagons. Due to the small number of spins, there are more pairwise interactions than higher-order couplings; however, practically useful examples will typically have more spins and therefore possess fewer pairwise interactions. (b-c) The dependence of the Kullback-Leibler divergence on the entropy ($H$) for various model p-spin systems (using a base two logarithm, $\text{log}_{2}$ as opposed to the more traditional natural log in Eqn.~\ref{eqn:shannon_entropy} and Eqn.~\ref{eqn:kullback_leibler}). b) Randomly generated 3-body interactions networks composed of $11\leq N_{q} \leq 20$ spins are mapped onto one- and two-body reduced models. c) Randomly generated 4-body interactions networks composed of $11\leq N_{q} \leq 22$ spins are mapped onto a pairwise model. The data are segregated according to a measure of the network ``interaction strength'' $\alpha$, which is proportional to the product of the standard deviation of the randomly generated spin couplings, $s$, and the (normalized) number of non-zero p-body couplings  i.e., $\alpha = sM_{J} / N_{q}$. Figure adapted from~\cite{pairwise_sufficiency}.}
\label{fig:pairwise_sufficiency}
\end{figure}

An ensemble of $p=3$ and $4$ models were randomly generated--possessing a range of entropies--by varying the number of spins ($N_{q}$), the number of non-zero p-body couplings ($M_{J}$), and the values of the couplings ($\boldsymbol{J}_{p}$). These higher-order models were mapped onto pairwise spin models ($p=2$), subject to the constraint that the two-body spin probabilities of the pairwise model matched the reduced two-body distributions obtained from the higher-order probability distribution. A schematic representation of this mapping for $p=3$ is shown in Fig.~\ref{fig:pairwise_sufficiency}a: the top network depicts a three-body spin model with four spins (circles), where all of the 3-body couplings (squares) are chosen to be non-zero. On the bottom is the corresponding pairwise spin model, where the 2-body couplings are represented as hexagons. 

In order to evaluate how well the reduced distributions replicated the targets, a measure of similarity between two distributions is required. Various metrics exist for this purpose, many of which have seen renewed interest for their utility in rigorous force-field optimization and systematic coarse-graining.~\cite{statistical_distances,relative_entropy,noid_perspective} For the p-spin study, the authors adopted the popular Kullback-Leibler divergence
\begin{equation} 
\label{eqn:kullback_leibler}
\begin{split}
& D_{\text{KL}}[P(\boldsymbol{X})\big|\big|P_{0}(\boldsymbol{X})]\equiv \sum_{\boldsymbol{X}}\big[P_{0}(\boldsymbol{X})\text{ln}P_{0}(\boldsymbol{X})-P_{0}(\boldsymbol{X})\text{ln}P(\boldsymbol{X})\big] \\
& = \langle \text{ln}P_{0}(\boldsymbol{X}) \rangle_{P_{0}(\boldsymbol{X})} - \langle \text{ln}P(\boldsymbol{X}) \rangle_{P_{0}(\boldsymbol{X})}
\end{split}
\end{equation} 
which is a measure of the degree to which $P(\boldsymbol{X})$ overlaps with $P_{0}(\boldsymbol{X})$.~\cite{relative_entropy} The KL-divergence is also known as the relative entropy in the field of bio-molecular coarse-graining.~\cite{relative_entropy} 

Adopting the higher-order model as $P_{0}(\boldsymbol{X})$, Fig.~\ref{fig:pairwise_sufficiency}b demonstrates that there are two regimes where the higher-order distributions are well-represented by pairwise models (i.e., $D_{\text{KL}} \approx 0$): when the entropy of the target is either very low or very high. For high entropy systems, this agreement is trivial: if there is relatively little order, there are not particularly stringent requirements on the (necessarily weak) couplings that can generate such a system. Even one-body, or independent, models work well when the entropy is sufficiently high, as can be seen in the right hand side of Fig.~\ref{fig:pairwise_sufficiency}b. However, more surprising is the observation that very ordered systems (i.e., $H \rightarrow 0$) are also well-represented by pairwise models. Furthermore, pairwise sufficiency was also found to be favored when the higher-order networks are densely interacting--as quantified by the parameter $\alpha$. In Fig.~\ref{fig:pairwise_sufficiency}c, the ensemble of spin models were first sorted by $\alpha$ of the target network and then plotted separately for different ranges in $\alpha$. Fig.~\ref{fig:pairwise_sufficiency}c shows a notable decrease in $D_{\text{KL}}$ as $\alpha$ increases.

Taken together, pairwise models are particularly well suited to represent higher-order networks when the latter has low entropy and is densely interacting. Low entropy systems necessarily possess correlations among constituent particles, spins, etc.; the lowest entropy spin system occurs when all spins move coherently (all up or all down), for instance. While such an arrangement may be obtained via many higher-order models, sufficiently strong positive pair interactions between all spins will achieve the same structure. Similarly, finite-sized correlated spin groups (i.e., clusters) can be obtained by employing strong pair interactions between the spins comprising each spin cluster while setting all other pair interactions to significantly smaller values. Interestingly, prior work on random constraint satisfaction anticipates the presence of such large correlated groups--amenable to representation via pair interactions--when the underlying distribution is densely interacting.~\cite{rcs_1,rcs_2} 

It has been suggested that biological systems can be regarded as evolutionarily derived solutions to complex constraint satisfaction problems~\cite{spins_evolution,pairwise_sufficiency}--ideal candidates for modeling with pairwise statistical models, according to the above analysis. Therefore, pairwise sufficiency may actually be characteristic to many biological systems--not just for the viral sequences reviewed earlier. Indeed, other successes in the modeling of biological processes with pairwise maximum entropy have been demonstrated.~\cite{spin_glasses_and_max_ent,max_ent_antibody,max_ent_ecoli,max_ent_flocks,bio_near_criticality} For example, such methods have been used to construct a model for gene expression levels under fluctuating cellular metabolic conditions.~\cite{gene_interactions} DNA micro-arrays can measure thousands of gene expression levels at a time, resulting in a high-dimensional data set. Similar to the viral sequences above, pairwise entropy maximization provides an avenue to both circumvent the curse of dimensionality and extract pair interactions between genes.

This approach was applied to gene expression data of the well-studied \emph{Saccharomyces cerevisiae} (i.e., Baker's yeast) under conditions which facilitate energy metabolic oscillations.~\cite{gene_oscillations} The resultant fluctuations in transcription levels yield an empirical ensemble of gene expression data. Micro-array data was encoded as an array of variables,  $\boldsymbol{G}\equiv\{g_{i}\}$, where $g_{i}$, is one of $m$ continuous, scalar, gene expression levels. Constraints imposed on the maximum entropy model include the empirically observed one and two-body gene expression correlations, $\langle g_{i} \rangle_{P_{0}(\boldsymbol{G})}$ and $\langle g_{i}g_{j} \rangle_{P_{0}(\boldsymbol{G})}$, respectively; this yields a model analogous to Eqn.~\ref{eqn:p_spin_model}, where both $p=1$ and $2$ contributions are present and the spins are continuous.

\begin{figure}[h]
\centering
\includegraphics[width=5.5in]{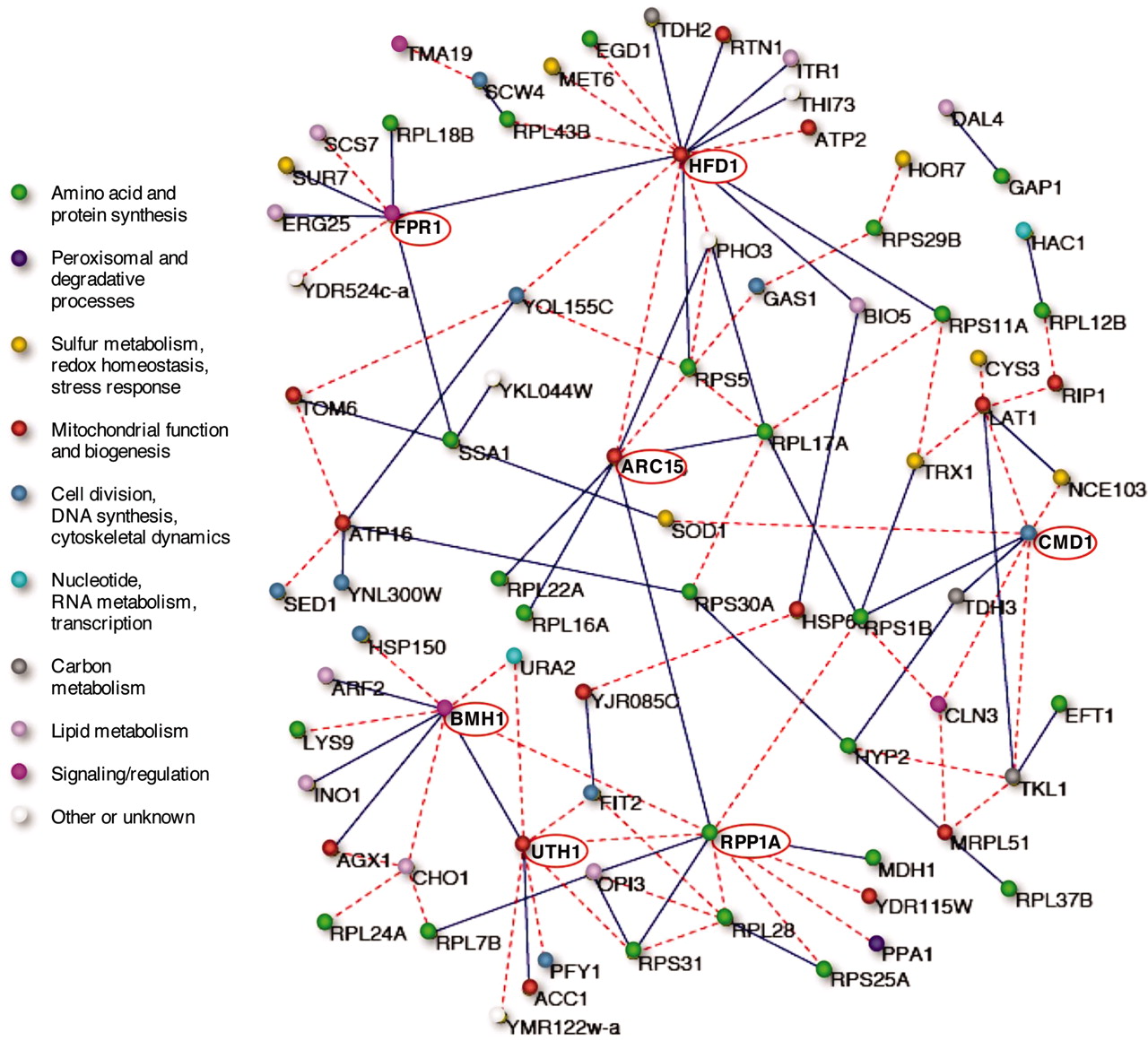}
\caption{Strongest 110 pair interactions (network connections) connecting genes (nodes) in the maximum entropy model. Each of the genes is labeled as well as color coded to indicate an approximate cellular function role. Solid-blue and dashed-red lines indicate positive and negative couplings, respectively, and the seven most-connected nodes (called ``hubs'') are circled. Figure reproduced from~\cite{gene_interactions}.}
\label{fig:gene_network}
\end{figure}

The benefit of using SI for gene expression data as opposed to a more common correlation-based analysis (e.g., clustering)~\cite{gene_clustering} is the construction of a network that quantifies the direct interactions between genes, possibly uncovering the driving force behind the observed correlations. Using correlation data alone only indicates which gene levels change together, but not what elements are driving those fluctuations. With the maximum entropy model though, it is easy to identify important governing genes, i.e., the nodes of high connectivity. For example, Fig.~\ref{fig:gene_network} has seven nodes with six or more connections, each of which corresponds to an important cellular protein for nutrient signaling. 

Furthermore, the model can uncover relationships between genes that are not part of the same cellular process. This can be seen visually in Fig.~\ref{fig:gene_network} by the many interactions between genes of different colors, where color indicates the cell process associated with a given gene. Therefore, while the protein expression patterns of \emph{S. cerevisiae} are relatively well understood and characterized, this need not be the case for the maximum entropy model to discover gene interactions and ``hubs'' that govern cellular activity and system dynamics within the cell. 

\section{Statistical mechanical inverse design of structure and properties}

The discovery of materials possessing a desired attribute is a long-standing design challenge that has been facilitated by computer simulation studies. Forward design--searching through parameter space to find favorable combinations that yield materials with some chosen characteristic--has benefited from recent advances in computational power.~\cite{janus,poly_packing,empty_liquids,lamellae,patchy_micelles} However, combinatorial space increases exponentially with dimensionality; therefore, such approaches can still be hampered by the cost associated with exploration of vast swathes of parameter space as well as potential couplings between tunable parameters. Inverse design (ID) methodologies--those characterized by an optimization framework typically used to iteratively update the parameters in a systematic fashion--represent an alternative approach. Inverse methods of statistical mechanics have facilitated the discovery of conditions (particle interactions, external fields, etc.) that lead to assembly of novel architectures.~\cite{jamming_design_3,nucleation_id,template_directed,st_inv_des_review,swarm_id}

One challenge to ID is to delineate an effective scheme to update the parameters; SI techniques can be leveraged to construct an optimization framework for such purposes. For instance, the strategy outlined in Sect.~\ref{sec:bio} for modelling an unknown probability distribution function can be adapted for application to complex materials design problems by ``mapping'' a contrived probability distribution ($P_{\text{tgt}}(\boldsymbol{X})$)--which has been intentionally imbued with some desired property--onto a statistical mechanical model. Standard SI approaches can then be used to discover the optimal parameters needed to reproduce the property present in $P_{\text{tgt}}(\boldsymbol{X})$. 

SI, therefore, can provide a prescription for performing updates to a probability distribution $P(\boldsymbol{X})$ that is constrained by a statistical mechanical model. In the context of equilibrium molecular simulation, the clear choice for the probability of observing a given configuration $\textbf{X}$ is the Boltzmann distribution

\begin{equation} \label{eqn:canonical_prob}
P(\textbf{X}|\boldsymbol{\theta})\equiv \dfrac{\exp[-\beta U(\textbf{X}|\boldsymbol{\theta})]}{Z(\boldsymbol{\theta})}, \text{    where  } Z(\boldsymbol{\theta})\equiv \sum_{\textbf{X}}^{} \exp[-\beta U(\textbf{X}|\boldsymbol{\theta})]
\end{equation} 
The probability for the configuration $\textbf{X}$, where $\textbf{X}$ can represent continuous particle coordinates, discrete spins, etc., can be tuned via the set of variables, $\boldsymbol{\theta}\equiv\{\theta_{i}\}$ which parametrize the configuration energy $U(\textbf{X}|\boldsymbol{\theta})$. The focus of the remainder of this section is primarily on two recent advances in novel materials discovery, each pertaining to complementary aspects of materials design: structure and properties. Both strategies optimize a probability distribution toward the design goal, where the optimization is guided by SI.

\subsection{Self-Assembly}
\label{subsec:self-assembly}

Top down fabrication of materials with desired nanoscale structural characteristics remains a formidable technological challenge.~\cite{tom_avni_perspective} Spontaneous self-assembly of constituent nanoscale particles provides an alternative bottom up approach for realizing materials with such designer morphologies; however, careful tuning of inter-particle interactions is required.~\cite{granick,patchy_review,janus,poly_packing,empty_liquids,lamellae,patchy_micelles} A recent series of studies outlined a statistical mechanical ID strategy for the discovery of such interactions.

For a particle-based system, the relevant coordinate ($\boldsymbol{X}$) is the set of D-dimensional Cartesian particle positions, $\boldsymbol{R}\equiv [\boldsymbol{r}_{1},\boldsymbol{r}_{2},...,\boldsymbol{r}_{N}]$. Prior to performing SI, a contrived target probability distribution over configuration space, $P_{\text{tgt}}(\textbf{R})$, that yields an ensemble of ``desirable'' configurations must be specified.~\cite{clusters_pores_crystals,crystals,pores,many_pores,clusters} Simulations constrained by either many-body interactions or by use of multiple components that enforce the desired structure have been successfully employed for this purpose. Then the KL-divergence (Eqn.~\ref{eqn:kullback_leibler}) between $P(\textbf{R}|\boldsymbol{\theta})$ and $P_{\text{tgt}}(\textbf{R})$ is minimized (where $P_{\text{tgt}}(\textbf{R})$ plays the role of $P_{0}(\textbf{X})$ from Sect.~\ref{sec:bio}). In Eqn.~\ref{eqn:kullback_leibler}, the only term that depends on the tunable parameters $\boldsymbol{\theta}$ is the second ``overlap'' contribution, $\langle\text{ln}P(\textbf{X}|\boldsymbol{\theta})\rangle_{P_{\text{tgt}}(\textbf{X})}$. 

Maximizing $\langle\text{ln}P(\textbf{R}|\boldsymbol{\theta})\rangle_{P_{\text{tgt}}(\textbf{R})}$ with respect to $\boldsymbol{\theta}$ has an intuitive probabilistic interpretation under fairly nonrestrictive assumptions. In particular, it is equivalent to maximizing the probability for the parametric model, $P(\textbf{R}|\boldsymbol{\theta})$, to sample an independently identically distributed set of configurations $[\textbf{R}_{1},\textbf{R}_{2},...,\textbf{R}_{M}]$ drawn from $P_{\text{tgt}}(\textbf{R})$ in the large sample limit ($M\rightarrow \infty$)--formally known as the maximum-likelihood approach of SI.~\cite{bayesian_reasoning,statistical_learning} 

Direct maximization of $\langle\text{ln}P(\textbf{R}|\boldsymbol{\theta})\rangle_{P_{\text{tgt}}(\textbf{R})}$ is intractable due to the presence of $Z(\boldsymbol{\theta})$ in Eqn.~\ref{eqn:canonical_prob}; however, iterative updates to $\boldsymbol{\theta}$ are easy to compute via natural gradient ascent.~\cite{clusters_pores_crystals,crystals} Under the constraint of isotropic pair interactions, i.e., $U(\boldsymbol{R}|\boldsymbol{\theta})\equiv\sum_{i<j=1}^{N}u(r_{i,j}|\boldsymbol{\theta})$ where $r_{i,j}\equiv|\boldsymbol{r}_{i}-\boldsymbol{r}_{j}|$, gradient ascent optimization in the canonical ensemble corresponds to
\begin{equation} \label{eqn:pair_update}
\boldsymbol{\theta} \rightarrow \boldsymbol{\theta} + \alpha \int_{0}^{\infty} dr r^{D-1} \big[g(r| \boldsymbol{\theta})-g_{\text{tgt}}(r)\big] \big[\beta\boldsymbol{\nabla}_{\boldsymbol{\theta}} u(r|\boldsymbol{\theta})\big]
\end{equation}
where $g(r)$ is the radial distribution function, $\alpha$ is the ``learning'' rate which is empirically set to guarantee stability of the optimization, and the definition of $Z(\boldsymbol{\theta})$ from Eqn.~\ref{eqn:canonical_prob} has been used. We note the analogy between the above approach to that outlined in Sect.~\ref{sec:bio}; in both, pair correlations (here, $g(r)$) are used as a proxy for the full probability distribution. In a sense, MD configurations represent a familiar example of sparsely sampled data: there are infinite possibilities for particle arrangements, and only some finite subset are sampled in an MD simulation. 

\begin{figure}[h]
\centering
\includegraphics[width=5.5in]{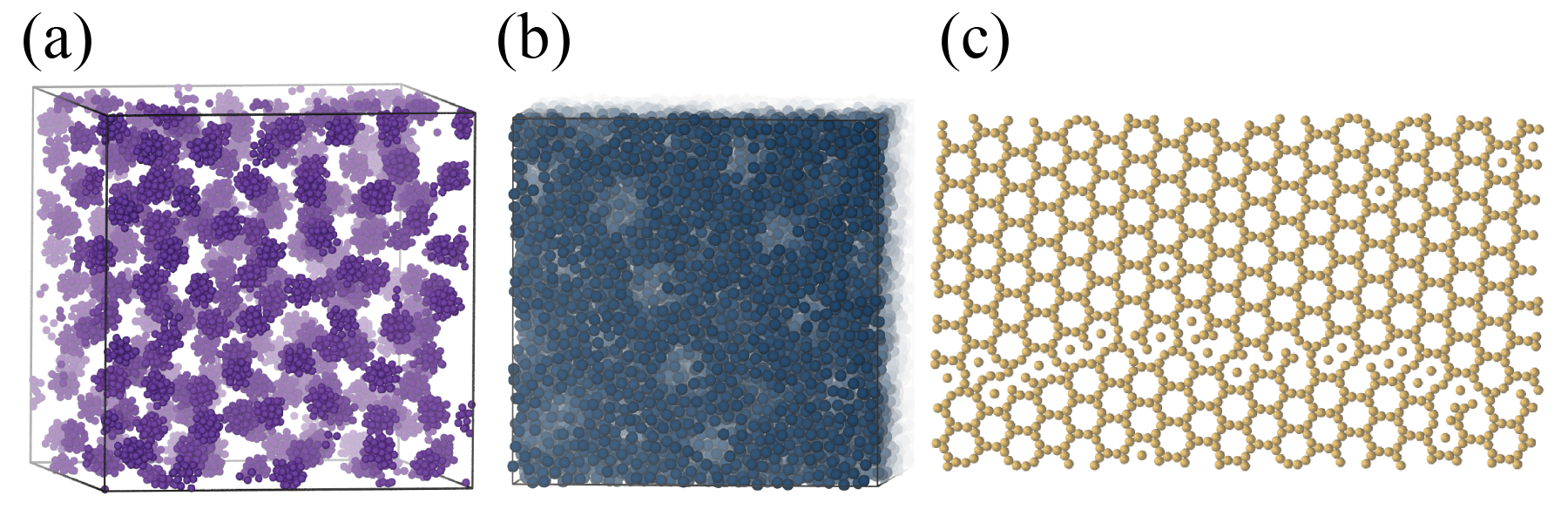}
\caption{Representative configurations obtained with the ID pair potentials for (a) clusters, (b) pores, and (c) the truncated hexagonal lattice.}
\label{fgr:ID_structures}
\end{figure}

The above approach has been successfully employed to engineer pair interactions that promote formation of clustered fluids, porous mesophase assemblies, and crystalline lattices; see Fig.~\ref{fgr:ID_structures}. The strategy is sufficiently general to handle all of these disparate architectures, with the primary difference being the construction of appropriate target ensembles. In all cases, standard simulation tools were employed with constraints chosen to enforce desired structural motifs. For cluster fluids, a many-body constraint upon the radius of gyration of each cluster was employed in conjunction with inter-cluster repulsive interactions that effectively enforced a minimum distance between particles in different clusters.~\cite{clusters,clusters_pores_crystals} In the case of a porous mesophase, a binary mixture of size-asymmetric hard-core-like particles was constructed, where the large particles created spherical zones of exclusion with respect to the smaller particles.~\cite{pores,clusters_pores_crystals} Finally, mutually non-interacting particles harmonically tethered to the desired lattice sites were used to emulate a crystalline lattice.~\cite{crystals,clusters_pores_crystals} 

Though some of the above architectures had previously been observed in MD simulations with pairwise interactions, particularly for clustered phases,~\cite{postulated_pore_phases_sear,simulated_pore_phases,generalized_cluster,bolhuis_clusters} no systematic framework existed for controlling the size of mesoscale structural features (i.e., clusters or the pores). By encoding the desired cluster or pore size in the target ensembles, interparticle interactions were discovered that reflected the prescribed mesophase dimension, allowing for the diameter of pores or the number of particles in a cluster to be tuned by modulating $P_{\text{tgt}}(\textbf{R})$. 

As can be visually gleaned from Fig.~\ref{fgr:ID_structures}, the desired motifs (clusters, pores, and crystals) all require highly structured particle arrangements, intuitively indicating a target ensemble with relatively low entropy. As discussed in Sect.~\ref{subsec:pairwise}, such low entropy ensembles are generally amenable to pairwise models, and the successes outlined above seem to support this finding.~\cite{pairwise_sufficiency} When there is flexibility in the creation of the target ensemble, a reasonable design principle might be to choose the lowest entropy ensemble that is practical. The systematic implementation of such a goal is an open question for future work.  

\subsection{Material Properties}

While Sect.~\ref{subsec:self-assembly} outlines a robust methodology for designing structure, the relationship between structure and material properties is a complicated many-body physics problem. Therefore, a systematic, statistical mechanical framework for the direct optimization of targeted material properties is complementary to the structure-based ID approach outlined above.

One such ID scheme leverages the probabilistic micro-state configuration information afforded by a statistical mechanics driven approach. The key requirement of the method is a link between configuration, $\textbf{R}$, and a configuration-dependent FOM, $F(\textbf{R})$, which quantifies the user's goal.~\cite{design_engines} For targeting a material property, $F(\textbf{R})$ could be the material property itself or a related metric. Derivation of a heuristic optimization strategy starts with an idealized case where $P(\textbf{R}|\boldsymbol{\theta})$ is infinitely flexible and can be additively updated by some $\delta P_{\text{id}}(\textbf{R}|\boldsymbol{\theta})$ at each individual $\textbf{R}$. While the choice of $\delta P_{\text{id}}(\textbf{R}|\boldsymbol{\theta})$ is non-unique, it is constrained by the requirement that normality of the probability is preserved, $\sum_{\textbf{R}}^{}\delta P_{\text{id}}(\textbf{R}|\boldsymbol{\theta})=0$, and that it generally increases (decreases) the probability for $\textbf{R}$ that has improved (decreased) fitness. One simple choice that guarantees both is
\begin{equation} \label{eqn:ideal_update}
\delta P_{\text{id}}(\textbf{R}|\boldsymbol{\theta})\propto P(\textbf{R}|\boldsymbol{\theta}) \big[F(\textbf{R}) - \langle F(\textbf{R}) \rangle_{P(\textbf{R}|\boldsymbol{\theta})}\big]
\end{equation}

Constraining the probability distribution to conform to a realistic statistical mechanical model necessarily limits the possibilities for $P(\textbf{R}|\boldsymbol{\theta})$; i.e., the probability of observing a single configuration cannot generally be tuned independently of the probabilities associated with other configurations. Thus, exactly realizing Eqn.~\ref{eqn:ideal_update} is not generally feasible. Instead, one seeks the best approximation to the ideal update afforded by a change in parameters ($\delta \boldsymbol{\theta}$) via
\begin{equation} \label{eqn:true_update}
\delta P(\textbf{R}|\boldsymbol{\theta})\equiv \delta \boldsymbol{\theta}^{T} \boldsymbol{\nabla}_{\boldsymbol{\theta}}P(\textbf{R}|\boldsymbol{\theta})
\end{equation}
by minimizing the average (over $P(\textbf{R}|\boldsymbol{\theta})$) sum of square differences between the ideal and realizable updates. Similar to Sect.~\ref{subsec:self-assembly}, the updates to the parameters can be computed iteratively.

\begin{figure}[h]
\centering
\includegraphics[width=5.5in]{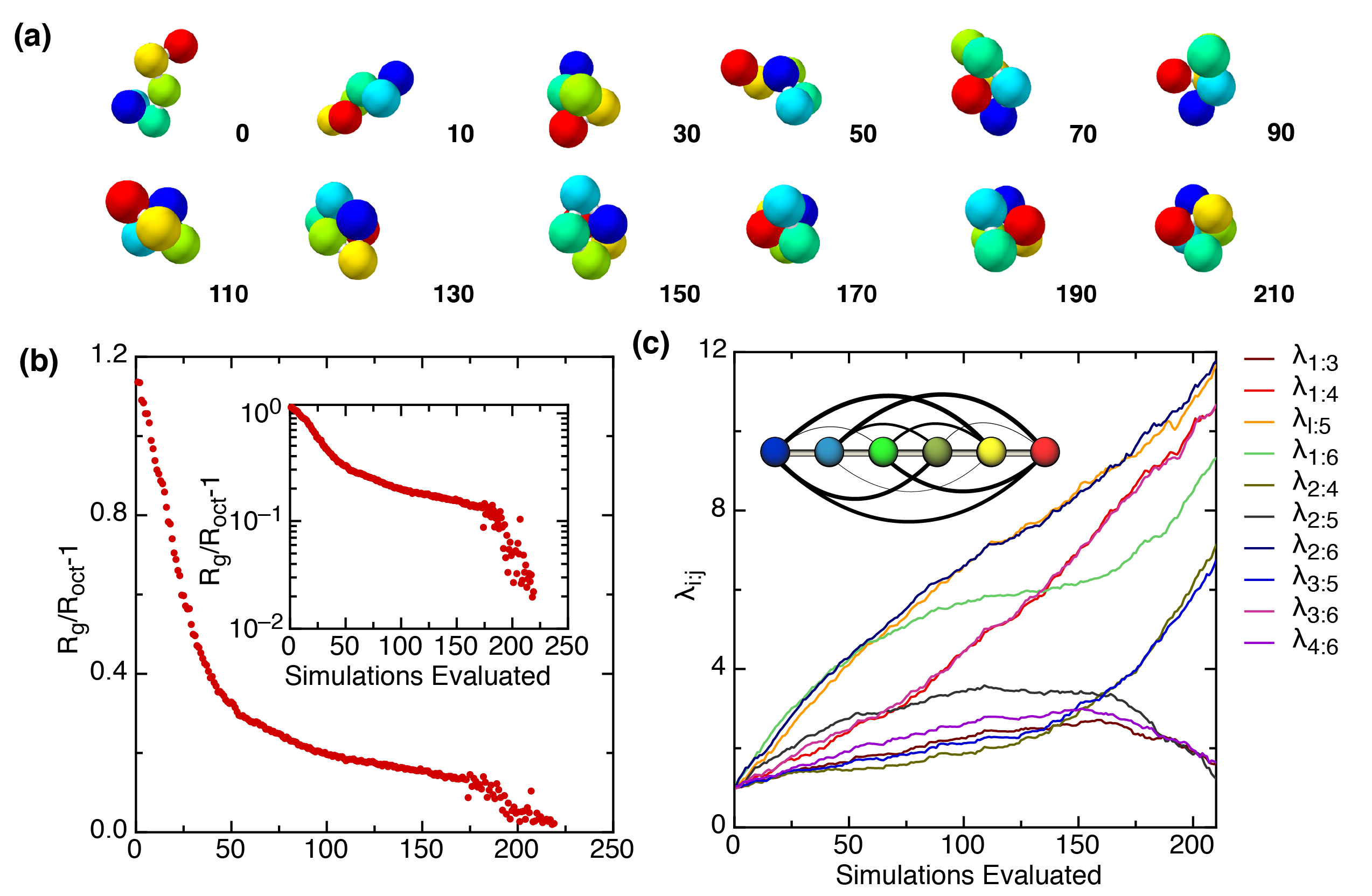}
\caption{ID of a polymer chain that folds into an octahedron. (a) Representative polymer configurations at specified optimization step numbers. (b) Convergence of the optimization as measured by the mean percent deviation between the average polymer radius of gyration ($R_{g}$) and the desired octahedron value ($R_{\text{oct}}$). (c) Evolution of the energetic coupling constants during the optimization. (\emph{inset}) Schematic representation of optimized interaction network, where thicker lines imply stronger couplings. Figure reproduced from~\cite{design_engines}.}
\label{fgr:polymer_octahedra}
\end{figure}

The above methodology has been shown to be successful in a variety of contexts, including for a square-lattice Ising model where interaction parameters were tuned to maximize magnetization, optimization of a trap for a thermalized particle constrained to a 2D substrate, self-assembly of block copolymers on a chemically patterned substrate, and discovery of pairwise interactions that result in a six-bead chain spontaneously folding into an octahedron. For the last case, each bead, a coarse-grained representation of a polymer, was modeled as a hard sphere connected by fixed bond lengths between nearest neighbors. Instead of optimizing for structure directly as in Sect.~\ref{subsec:self-assembly}, the radius of gyration [$R_{g}(\boldsymbol{R})$] of the chain was used to quantify the polymer shape. The qualitative progress of the optimization, progressing from an extended chain to a compact object, can be seen in Fig.~\ref{fgr:polymer_octahedra}a, where the index indicates the optimization step. More quantitatively, the metric $R_{g}/R_{\text{oct}}-1$ (percent deviation from target, essentially an inverse fitness) decreased monotonically over the optimization; see Fig.~\ref{fgr:polymer_octahedra}b, where $R_{g}\equiv \langle R_{g}(\boldsymbol{R})\rangle_{P(\textbf{R}|\boldsymbol{\theta})}$ is the mean radius of gyration. After roughly 200 iterations, the mean deviation from an octahedral $R_{g}$ was only around 1\%; it is visually apparent in Fig.~\ref{fgr:polymer_octahedra}a that the final steps in the optimization resulted in an octahedron.

The evolution of the couplings ($\boldsymbol{\lambda}$) over the course of the optimization are shown in Fig.~\ref{fgr:polymer_octahedra}c. The inset schematic shown in Fig.~\ref{fgr:polymer_octahedra}c summarizes the optimized results as the interaction network where thicker lines indicate stronger couplings. Though symmetry is not explicitly enforced, the mirror symmetry present in both the bead model and the octahedron is manifest (to a good approximation) in the interactions, i.e., $\lambda_{1:5} \approx \lambda_{2:6}$, etc. Furthermore, the resulting interactions are relatively intuitive; for instance, pairs of beads that are positioned co-linearly with the center of the octahedron, and therefore are farther apart, have the weakest couplings ($\lambda_{1:3}, \lambda_{2:5}, \lambda_{4:6}$).      

For the above proof-of-concept example, the relationship between structure and property ($R_{g}$) is rather trivial, but this need not be the case. In principle, the method can be straightforwardly applied for \emph{any} material property which uniquely depends on particle coordinates. For example, other structural metrics like the fractal dimension and average spacing between particles are feasible. More generally, quantities that use particle coordinates in intermediate theoretical calculations, such as electron hopping conductivity,~\cite{conductivity} can be optimized.  

\section{Conclusions and outlook}

Techniques of ML and SI are useful in a wide variety of physical chemistry applications in both experimental and computational contexts--a sampling of which has been provided in this review. For example, Bayesian optimization is straightforward to incorporate into experimental design loops to systematize the selection of new samples to prepare and measure. On the computational side, ML tools have been incorporated into MD simulations to accelerate QM force calculations, though other expensive computations for simulation could potentially be facilitated by ML as well. Also reviewed were applications of SI for modeling complex systems, with case studies taken from 1) biology in the form of viral sequences and gene expression data and 2) materials design of both structure and properties.

It is relatively early on in the adoption of ML and SI tools in the physical sciences. Undoubtedly many new applications remain to be discovered. However, one broad implication of further integrating ML and SI into research is the promise of increased opportunities for cooperation between experimental and computational research. For example, the viral sequence study used real clinical data to create an empirical computational model that may be translated into rational, real world, vaccine design. Similarly, Bayesian optimization can leverage theoretical frameworks to construct features to be input into a numerical optimization procedure that guides experimental design. In the future, perhaps the primary utility of ML and SI methodologies will be the erosion of the distinctions between experiment, computation, and theory, ultimately facilitating both collaboration and discovery.





\section*{Acknowledgments}
We are grateful for comments and suggestions provided by Andrew Ferguson on a working draft of this review.
Work from the authors presented here was partially supported by the National Science Foundation (1247945) and the Welch Foundation (F-1696). We acknowledge the Texas Advanced Computing Center (TACC) at The University of Texas at Austin for providing HPC resources.




\end{document}